\documentclass[12pt]{iopart}
\usepackage{amsfonts}
\begin{document}
\title[SWOF and gravitational energy quasilocalization]{Sen-Witten orthonormal three-frame and gravitational energy
quasilocalization}
\author{ V Pelykh}
\address{ Pidstryhach Institute of Applied Problems in Mechanics and
Mathematics\\ Ukrainian National Academy of Sciences, 3B Naukova
Str.,\\ Lviv, 79601, Ukraine}

\ead{pelykh@lms.lviv.ua}

\begin{abstract}
 Expression for the Witten-Nester 4-spinor 3-form
of the Hamiltonian density of  gravitational field in the asymptotically flat
space-time in terms of the Sommers-Sen spinors, direct with a certain
orthonormal  three-frame connect, is obtained. A direct connection between
the one  and the ADM Hamiltonian density in the Sen-Witten frame is
established on this basis.
\end{abstract}

\submitto{\CQG}

\maketitle
\section {Introduction}

The equivalence principle excludes a possibility for existence of the
gravitational energy density,  however, there is possible its
quasilocalization in the Penrose conception \cite{pen1}. This conception is
realized in several proposals for the quasilocal energy-momentum [2-7].
According to the Nester and coauthors approach, for each gravitational energy
momentum pseudotensor there is  Hamiltonian boundary term, and the
energy-momentum in a domain, bounded by close 2-surface, depends on the field
values and the frame of reference on the 2-surface. Various criteria are
insufficient and, most probably \cite{nes40}, will be insufficient for
selecting a unique Hamiltonian boundary term. Variety of these terms is
characterized by different choices of dynamic variables (metric, orthonormal
frame, spinors), boundary conditions and reference configurations. According
to this there appears a problem of the   different  Hamiltonians comparing
\cite{nestu, nestu30}. Among criteria that must be satisfied by the
quasilocal energy-momentum density, at least in asymptotically Minkowskian
space \cite{chr}, must be positivity. It can be ensured by finding the
locally non-negative Hamiltonian density dependent on the Sen-Witten spinor
according to Witten \cite{wit}, or by applying the ADM Hamiltonian and the
Nester special orthonormal frame (SOF) \cite{nesPL}.

 In the asymptotically flat space the Hamiltonian is of the general form
\cite{isnes}

\begin{equation}
  \label{1}H\left( {N} \right) = {\int\limits_{\Sigma} {\left( {N{\cal H} +
N^{a}{\cal H}_{a}} \right)}} dV + {\oint\limits_{\partial \Sigma}  {B}} $$
\end{equation}
and includes the Regge-Teitelboim boundary term \cite{regge} at spacial
infinity.

Grounding and developing the Wittenian proof of the positive energy theorem
(PET), Nester \cite{nesLec} proposed an expression for the Hamiltonian
density as the 4-covariant quadratic spinor 3-form:
\begin{equation}
\label{2} {\cal H}\left( {\psi}  \right): = 2{\left[ {D\left( {\overline
{\psi}\land  \gamma _{5} \gamma}  \right)\land D\psi - D\overline {\psi}
\land D\left( {\gamma _{5} \gamma \psi} \right)} \right]}
\end{equation}
where
\[D\psi = \rmd\psi + {\frac{{1}}{{2}}}\omega ^{\mu \nu} \sigma
_{\mu \nu}  \psi {\rm ,} \quad \sigma _{\mu \nu}  = {\frac{{1}}{{2}}}{\left[
{\gamma _{\mu}  ,\gamma _{\nu }} \right]}, \quad \gamma = \gamma _{\mu}
\theta ^{\mu}\]
 \[ \quad
\gamma _{\mu}  \gamma _{\nu}  + \gamma _{\nu} \gamma _{\nu}  =
2g_{\mu \nu } ,\qquad  \gamma _{5}^{2} = - E, \quad \gamma _{5} =
\gamma ^{0}\gamma ^{1}\gamma ^{2}\gamma^3 .
\]

From expression (\ref{2}) one can obtain the following expression
for ${\cal H} \left({\psi} \right)$:

\begin{equation}
\label{3} {\cal H}\left( {\psi}\right) = 4D\overline {\psi} \land\gamma _{5}
\gamma \land D\psi = 4\nabla _{\pi} \overline {\psi} \left( {\gamma ^{\mu}
\sigma ^{\pi \rho} + \sigma ^{\pi \rho} \gamma ^{\mu} } \right)\nabla _{\rho}
\psi d\Sigma _{\mu}
\end{equation}
where $d\Sigma _{\mu}  = {\frac{{1}}{{3!}}}\sqrt {{\left| {g} \right|}}
\varepsilon _{\mu \nu \pi \sigma}  dx^{\nu} \land dx^{\pi}\land dx^{\sigma}
$.

In the Gaussian normal system of coordinates in  the neighborhood of
arbitrary spacelike hypersurface $\Sigma $, the Hamiltonian density (\ref{3})
can be written as a sum of positive and negative definite  components
\cite{nesLec}

\begin{equation}
\label{4} {\cal H}\left( {\psi}  \right) = - 4g^{ab} D_{a } \psi ^{ +} D_{b}
\psi + D_{a} \overline {\psi}  \left( {\gamma ^{d} \gamma ^{a} \gamma ^{b} +
\gamma ^{a} \gamma ^{b} \gamma ^{d} } \right)D_{b}  \psi\, d\Sigma _{0}
\end{equation}
and be locally non-negative if $SL(4, \mathbb C)$-spinor $\psi $ on the
spacelike hypersurface $\Sigma $ satisfies the Sen-Witten equation (SWE)

\begin{equation}
\label{5} \gamma ^{a} D_{a}  \psi = 0{\rm .}
\end{equation}
Expression (\ref{4}) cannot give the true positive energy density
for the gravitational field because $\psi $, as solution of the
SWE, is a nonlocal functional on the initial data $(h,K,\Sigma)$
set; therefore, a concept of the locally non-negative density of
the gravitation energy is treated as the locally non-negative
functional on the set of initial data $(h,K,\Sigma)$ and the
boundary values of function $\psi $. The gravitational Hamiltonian
density (\ref{4}) has significant advantages in comparison with
the other ones: except a fact that it is explicitly 4-covariant,
the gravitational Hamiltonian, which includes it, allows to prove
that  the total 4-momentum and the Bondi 4-momentum are timelike.
To its liabilities Nester and Tung had referred the physical
mysteriousness of the Sen-Witten spinor field, and absence of the
direct relation to the Hamiltonian density in the SOF method
\cite{nesjmp, nesPL, nesCQG}. For establishing such relation,
Nester and Tung \cite{nestu} had developed a new method of proving
the PET and the gravitational energy localization, which employees
the 3-dimensional spinors and a new identity connecting the
3-dimensional scalar curvature to the spinor expression in the
Hamiltonian. The Einstein 3-spinor Hamiltonian with a zero shift
the authors obtained in the form \footnote{ In this formula and in
some next formulas we change the signs, comparing with the
original papers, according to the chosen here convention that a
signature is $(+---)$.}
\begin{eqnarray}
\label{6}\fl H = {\int\limits_{\Sigma}  {{\left[ {\varphi ^{ +} \varphi g^{ -
{\frac{{1}}{{2}}}}\left( { \pi^{ab} \pi_{ab} -
{\frac{{1}}{{2}}}\pi^{2}}\right)
 - 4\left( {g^{a
b} \nabla _{a } \varphi ^{ +} \nabla _{b}  \varphi + \nabla _{a} \varphi ^{ +
}\sigma ^{a} \sigma ^{b} \nabla _{b} \varphi} \right)} \right]}}} d^{3}x
\end{eqnarray}
from which  follows a conclusion that the density is non-negative definite,
if on the maximal hypersurface the asymptotically constant spinor $\varphi $
satisfies the Dirac equation in the 3-dimensional space

\begin{equation}
\label{7} \sigma ^{a}\nabla _{a} \varphi = 0{\rm .}
\end{equation}

The main result of the  Nester, Tung \cite{nestu} and the Nester, Tung,
Zhytnikov \cite{nestuzhy} works is formulated in the form of a statement that
between the localization method, based on the 4-covariant spinor Hamiltonian,
and the SOF-based method there exists a close connection owing to the
3-spinor Hamiltonian (\ref{6}).

Such statement is grounded on the two circumstances: 1) among terms of which
the 4-covariant spinor density consists, the 3-spinorial density is present;
2) between the 3-spinor field variables there exists, as Nester and Tung
declared, a close relation, since from the 3-dimensional Dirac equation
(\ref{7}) it follows that
\begin{equation}
\label{8} \sigma ^{a} \nabla _{a}  \varphi = \sigma ^{a} \varphi _{,a} -
{\frac{{1}}{{2}}}\tilde {q}_{b} \sigma ^{b} \varphi + {\frac{{1}}{{4}}}{\rm
i}*q\varphi = 0
\end{equation}
where forms  $q$ and $\tilde {q}$ are defined in the following
way:

\begin{equation}
\label{9} q = \theta _{\hat{a}}  \land d\theta ^{\hat{a}} {\rm ,}
\quad \tilde {q} = i_{\hat{a}} d\theta ^{\hat{a}}
\end{equation}
and fix SOF on the asymptotically flat surfaces by means of the Nester gauge
\begin{equation}\label{10}
*q=0, \quad \tilde{q} ={\rm d}\ln\Phi,
\end{equation}
 where $\Phi$ is arbitrary everywhere positive function.
  Nester, Tung and Zhytnikov results do not connect
the
 Dirac equation itself and the Nester gauge by the equivalence
 relationship of a certain type, and do not establish the explicit and unique
 connection in all points of $\Sigma$ (see, for example, \cite{dim1,dim2,nesCQG}) between
the variables of the 3-spinor field and the SOF variables. That is why a
 search for the
 valuable  grounding of a statement about existence of a
close correlation between both approaches remains topical.

 We  propose further a new insight on the problems of
this correlation that is fully correct for the case of maximal
hypersurface and is grounded asymptotically   in the case of quite
arbitrary hypersurfaces. The reason is known: the linear equations
for the spinor fields become nonlinear after transition to the
respective tensor functions.

\section{Direct link between the 4-covariant spinor 3-form and the
Einstein Hamiltonian}

 In \cite{peljpa} we have proved the two
 theorems:

{\bf Theorem 3.} {\it Let:\\
 a) initial data set $(h,K,\Sigma)$ be
asymptotically flat by Reula \cite{reu};\\\
b) everywhere on $\Sigma$ the matrix of the spinorial tensor
\begin{equation}\label{11}
C_A{}^B:=\frac{\sqrt2}{4}D_{A}{}^{B}{\cal K}+\frac14 \varepsilon
_A{}^B\left(2{\cal K}^2+\frac12{\cal K}_{\pi\rho}{\cal
K}^{\pi\rho}+\mu\right)
\end{equation}

\hspace{-5mm} has, at least, one non-negative eigenvalue, for
definiteness $C_0$;\\ c) ${\rm Re}\,\lambda^0_{\infty}$ or ${\rm
Im}\,\lambda^0_{\infty}$ asymptotically nowhere are equal to
zero.

Then the asymptotically constant nontrivial solution $\lambda^C$ to the SWE
does not have the knot points on $\Sigma_t$.}

Here $\lambda^A_{\infty}$ --- asymptotic value of the $SU(2)$-spinor field
$\lambda^A$, which is a result of the Sommers and Sen reducing \cite{som,
sen} the $SL(2,\mathbb C)$ structure to the $SU(2)$ structure on the
spacelike hypersurface $\Sigma$ with the unit normal one-form $n$.

{\bf Theorem 4.} {\it   Let the conditions of Theorem 3 be fulfilled. Then
everywhere on $\Sigma$ there exists a two-to-one correspondence between the
Sen-Witten spinor field $\lambda^A$ and the Sen-Witten orthonormal frame
$\theta^a$.}

Here the Sen-Witten orthonormal frame (SWOF) we call a 3-frame
$\theta^{\hat{a}}$ distinguished by the gauge conditions
\begin{eqnarray}
\varepsilon^{\hat{a}\hat{b}\hat{c}}\omega_{\hat{a}\hat{b}\hat{c}}
\equiv*q=0, \qquad\omega^{\hat{a}}{}_{\hat{1}\hat{a}}\equiv
-\widetilde q _{\hat{1}}= -F_{\hat{1}},\qquad
\omega^{\hat{a}}{}_{\hat{2}\hat{a}}=
-\widetilde {q}_{\hat{2}}=-F_{\hat{2}},\nonumber\\
\label{12} \omega^{\hat{a}}{}_{\hat{3}\hat{a}} =-\widetilde
q_{\hat{3}}=-{\cal K}-F_{\hat{3}},
 \end{eqnarray}
 where $\omega_{\hat{a}\hat{b}\hat{c}}$ are the
connection one forms coefficients,  $F={\rm d}\ln\lambda$, and
$\lambda=\lambda_A\lambda^{A+}$;  "hats" distinguish orthonormal
frame indices.

The SWOF   generalizes the Nester SOF because gauge (\ref{10}) can be written
as
\begin{equation}\label{13}
*q:=\varepsilon^{\hat{a}\hat{b}\hat{c}}\omega_{\hat{a}\hat{b}\hat{c}}=0,
\qquad\widetilde q
_{\hat{b}}:=-\omega^{\hat{a}}{}_{\hat{b}\hat{a}}=\partial_{\hat{b}}\ln\Phi.
\end{equation}

The above mentioned correspondence between the Sen-Witten spinor and  SWOF is
of the form:
\begin{equation} \label{14}
\theta^{\hat{1}}=\frac{\sqrt2}{2\lambda}(L+\overline L),\quad
\theta^{\hat{2}}=\frac{\sqrt2}{2\lambda \rmi}(L-\overline L),
\quad\theta^{\hat{3}}=\widetilde L
 \end{equation}
  where $ L=-\lambda_A\,\lambda_B$\, and $\widetilde L=\mid
L\mid^{-1}*\left(L\land\overline L\right)$ is the nonzero spatial
one--form.

 In other words, we can  say that if the spinor field
$\lambda_A$ satisfies the SWE  and  conditions of Theorem 3, then everywhere
on $\Sigma_t$ exists the orthonormal 3-frame which satisfies conditions
(\ref{12}), and conversely.

 Frauendiener \cite {fra}  represented "squared neutrino
 equation",
obtained by Sommers \cite{som} from Weyl equation by means of
transformation (\ref{14}), as a conditions for a set of three
mutually orthogonal fields of equal length on $\Sigma$. Our
conditions (\ref{12}) are the other form of "squared zero-modes
neutrino equation" and the Frauendiener conditions. Such form of
these conditions will permit us further to apply them efficiently
for transformation of the ADM Hamiltonian along Nester's line. More
important is the following consideration. As it was pointed  by
Dimakis and M${\rm {\ddot u}}$ller-Hoissen \cite{dim1,dim2} (see
also \cite{ash}), the spinorial field as a solution to elliptic
equation may have zeros (knot points). Respectively, the Sommers
transformation does not exist on the knot surfaces, lines and points on
$\Sigma$; on such submanifolds "squared zero-modes neutrino
equation" and the Frauendiener conditions will be not satisfied,
and any correspondence between the Sen-Witten spinor field and the
SWOF will not exist. Our Theorem 4 solves this problem,
establishing the conditions for existence of transformation
(\ref{14}) everywhere on $\Sigma$.

Taking into account that the Hamiltonian density (\ref{2}) and the  SWE
were obtained by the spinor parameterization for the Hamiltonian
displacement, we write in terms of the Sommers-Sen spinors
\begin{equation}
\label{14+1}
N^\mu=\lambda^A\lambda^{\dot{A}}=\lambda^{(A}\lambda^{B)+}+N^\mu
n_\mu n^{AB}=
\lambda^{(A}\lambda^{B)+}+\frac1{\sqrt2}\lambda_D\lambda^{D+}
\varepsilon^{AB}.
\end{equation}
That is why $N\equiv N^0=\lambda_A\lambda^{A+}$ and $F={\rm d}\ln
N$. Note, that the Nester SOF approach does not limit a choice of
the dependence $N=N(\Phi)$ \cite{nesCQG}.

 We will further give  the 4-covariant Hamiltonian
density in terms of the Sen-Witten spinor using the SWE  in the form

\begin{equation}\label{15}
{\cal D}^B{}_C\lambda^C=0.
\end{equation}
An action of  operator ${\cal D}_{AB}$ on the spinor fields is
\[ {\cal D}_{AB}\lambda_C =D_{AB}\lambda_C+\frac{\sqrt2}{2}{\cal
K}_{ABC}{}^D\lambda_D \] where $D_{AB}$ --- the spinorial form of the
derivative operator $D_\alpha$ compatible with  metric $h_{\mu\nu}$ on the
$C^\infty$ hypersurface $\Sigma_t,\;\; {\cal K}_{ABCD}$ --- the spinorial
 tensor of the extrinsic curvature of hypersurface $\Sigma$.

The standard substitution transforms (\ref{3}) to the form
\begin{eqnarray}\fl {\cal H}(\varphi,\chi) = \left[- 2\sqrt {2}
 \left(
{n_{A\dot {A}} D_{\mu}  \varphi ^{A}D^{\mu} \overline {\varphi}
^{\dot {A}} + n_{A\dot {A}} D_{\mu}  \chi ^{A}D^{\mu} \overline
{\chi} ^{\dot {A}}}\right)\right.\nonumber
 \\ \label{16} \left.\lo+
2\left( {n^{B\dot {C}}D_{B\dot {A}} \overline {\varphi} ^{\dot
{A}}D^{\mu} \varphi ^{A} + n^{B\dot {C}}D_{B\dot {A}} \overline
{\chi}  ^{\dot {A}}D^{\mu} \chi ^{A}} \right)\right]d^{3}\Sigma.
\end{eqnarray}

Let us take into account that
\begin{eqnarray}\fl
 h^{\mu \nu} n_{A\dot {A}} D_{\mu}  \varphi ^{A}D_{\nu}  \overline {\varphi
} ^{\dot {A}} = \varepsilon ^{\dot {B}\dot {D}}\varepsilon
^{BD}n_{A\dot {A}} \left( {D_{B\dot {B}} \varphi ^{A}}
\right)\left( {D_{D\dot {D}} \overline {\varphi}  ^{\dot {A}}}
\right)\nonumber \\ \label{17}
 =2n^{R\dot {B}}n_{R}{} ^{\dot {D}}\varepsilon ^{BD}\left( {D_{B\dot {B}}
\varphi ^{A}} \right)\left( {D_{D\dot {D}} \overline {\varphi}
^{\dot {A}}} \right) = \left( {{\cal D}_{B}{} ^{R}\varphi ^{A}}
\right)\left( {{\cal D} ^{B}{}_{R}\overline {\varphi}  ^{\dot
{A}}} \right)n_{A\dot {A}},
 \end{eqnarray}
 and

\begin{equation}
\label{18} \left( {{\cal D}_{R}{} ^{B}\overline {\varphi}  ^{\dot
{A}}} \right)n_{A\dot {A}} = {\frac{{1}}{{\sqrt {2}} }}{\left[ { -
{\cal D}^{BR}\varphi ^{ +} _{A} - \sqrt {2} \left( {{\cal
D}^{BR}n_{A\dot {A}}} \right)\overline {\varphi}  ^{\dot {A}}}
\right]},
\end{equation}

\begin{equation}
\label{19} {\cal D}^{BR}n_{A\dot {A}} = {\cal K}^{BR}{}_{A\dot {A}} +
{\frac{{\sqrt {2}} }{{2}}}F_{A\dot {A}} \varepsilon ^{(BR)}= {\cal
K}^{BR}{}_{A\dot {A}}.
\end{equation}

Then
 \begin{eqnarray}\fl
 n_{A\dot {A}} h^{\mu \nu} \left( {D_{\mu}  \varphi ^{A}D_{\nu} \overline
{\varphi}  ^{\dot {A}} + D_{\mu}  \chi ^{A}D_{\nu} \overline
{\chi} ^{\dot {A}}} \right)
 = \left( {{\cal D}_{BR}\varphi _{A}}
\right)\left( {{\frac{{\sqrt {2} }}{{2}}}{\cal D}^{BR}\varphi ^{
+A}  + {\cal K}^{BRA}_{S} \varphi ^{ + S}} \right)\nonumber \\
\label{20}
 + \left( {{\cal D}^{BR}\chi ^{A}} \right)\left( {{\frac{{\sqrt {2}
}}{{2}}}{\cal D}_{BR}\chi^+ _{ A} + {\cal K}^{BRA}_{S} \chi ^{ + S}} \right).
 \end{eqnarray}
For transformation of the other terms we will use the identity
\begin{eqnarray} \fl
 D_{B\dot {A}} \overline {\varphi}  ^{\dot {A}} = D_{B\dot {A}} \left(
{2n^{\dot A {C}}n_{C\dot {C}} \overline {\varphi}  ^{\dot {C}}}
\right) = - {\frac{{2}}{{\sqrt {2}} }}D_{B\dot {A}} \left(
{n^{\dot A {C}}\varphi ^{ + }_{C}}  \right)
 =
\nonumber
\\
\label{21}
 - {\frac{{2}}{{\sqrt {2}} }}\left( {{\cal K}_{B\dot {A}}{} ^{\dot {A}C}\varphi ^{ +
}_{C} + {\frac{{1}}{{\sqrt {2}} }}{\cal D}_{B}{} ^{C}\varphi ^{
+} _{C}} \right)
 =
 - {\frac{{2}}{{\sqrt {2}} }}\left( {{\cal K}_{B}{} ^{C}\varphi ^{ +} _{C} +
{\frac{{1}}{{\sqrt {2}} }}{\cal D}_{B}{} ^{C}\varphi ^{ +} _{C}}
\right),
 \end{eqnarray}
and, therefore,
\begin{equation}
\label{22} n^{B\dot {C}}D_{B\dot {A}} \overline {\varphi} ^{\dot {A}}
D_{A\dot C} \varphi ^{A} = {\frac{{\sqrt {2}} }{{2}}}\left( {{\cal D}_{A}{}
^{B}\varphi ^{A}} \right)\left( {{\cal D}_{BC} \varphi ^{ +C}} \right) -
{\frac{{1}}{{2}}}{\cal K}_{BC}\varphi ^{ +C} {\cal D}_{A}{} ^{B}\varphi
^{A}{\rm .}
\end{equation}

The final expression for ${\cal H}(\varphi,\chi)$ we will give in
the form
 \begin{eqnarray}\fl {\cal
H}\left( {\varphi ,\chi}  \right) = \sqrt {2}\{ \left( {{\cal D}_{BR} \varphi
_{A}} \right)\left( {\sqrt {2} {\cal D}^{BR}\varphi ^{ + A} + {\cal
K}^{BRA}_{S}
\varphi ^{ + S}} \right)\nonumber \\
   +\left( {{\cal D}_{A}{} ^{B}\varphi^A}
\right)\left( { - {\cal D}_{BC} \varphi ^{ + C} + {\cal K}_{BC} \varphi ^{ +
C}} \right)\nonumber \\ +  \left( {{\cal D}_{BR} \chi _{A}} \right)\left(
{\sqrt {2} {\cal D}^{BR}\chi ^{ + A} + {\cal K}^{BRA}_{S} \chi ^{ + S}}
\right)
 \nonumber\\ \label{23}\lo+  \left(
{{\cal D}_{A}{} ^{B}\chi^{A}}  \right)\left( { - {\cal D}_{BC} \chi ^{ + C} +
{\cal K}_{BC} \chi ^{ + C}} \right)\}dV.
 \end{eqnarray}

The Hamiltonian 3-form ${\cal H}\left( {\varphi ,\chi}  \right)$ (\ref{23})
in comparison with the Hamiltonian 3-form,  obtained by Ashtekar and Horowitz
\cite{ash}, contains the terms with  the external curvature of hypersurface
$\Sigma $.

The first and the second terms are positive definite, and the next ones turn
to zero if on hypersurface $\Sigma $ the spinor fields $\varphi ^{A}$ and
$\chi ^{A}$  satisfy the SWE  (\ref{15}) .

On the other hand, the ADM Hamiltonian density, parameterized with
orthonormal 3-frames $\theta^{\hat{a}}$, is of the form
\cite{nesPL,nestu}
\begin{eqnarray}\fl
{\cal H}\left( {N} \right) = -2|h|^{{\frac{{1}}{{2}}}}\widetilde
q^a
\partial _a  N + N|h|^{{\frac{{1}}{{2}}}}\left( {{\cal K}^{a
b} {\cal K}_{ab}  - {\cal K}^{2}} \right)
-2|h|^{{\frac{{1}}{{2}}}}\left({\cal K}^a{}_b-\delta^a{}_B{\cal
K}\right)D_aN^b  \nonumber\\ + N|h|^{{\frac{{1}}{{2}}}}\left[
{q^{ab} q_{ab} + {\frac{{1}}{{2}}}\widetilde q^{a} \widetilde
q_{a}  - {\frac{{1}}{{6}}}(*q)^{2}} \label{24} \right]
\end{eqnarray}
 where the symmetric tensor $q_{ab}$,  vector
$\widetilde q_a$, and scalar $*q$ are defined by irreducible
decomposition
\[
C ^{a}{} _{bc}  = q^{ad} \varepsilon _{dcb} + {\frac{{1}}{{2}}}\left( {\delta
^{a} _{c} \widetilde q_{b}  - \delta ^{a} _{b}\widetilde q_{c} } \right) +
{\frac{{1}}{{3}}}*q\varepsilon ^{a}{} _{cb}.
\]
 Varying the lapse in (\ref{24}), we obtain the super-Hamiltonian
constraint in the form
\begin{equation}
\label{25}
 \fl 2\partial_k\left(|h|^\frac12q^k\right)+\frac12|h|^{\frac12}q^kq_k+
 |h|^\frac12\left[
{\cal K}^{mn}{\cal K}_{mn}-{\cal K}^2
+q^{mn}q_{mn}+\frac12\tilde{q}^a\tilde{q}_a-\frac16(*q)^2\right]=0.
\end{equation}

If the spinor fields $\varphi^A$ and $\chi^A$ satisfy the SWE and
conditions of Theorem 3, then  condition (\ref{12}) be fulfilled,
and vice versa . Then, on the one hand, ${\cal H}\left( {\varphi
,\chi}  \right)$ will be positive, and, on the other hand, this
will permit us to write ${\mathcal H}(N)$ in the SWOF, under the
necessary in this context limitation for $F$, and at $N^a=0$, in
the form
\begin{eqnarray}\fl
 {\cal H}^{SWOF}\left( {N} \right) = N|h|^\frac12\left({ {-\frac{{3}}{{2}}}}
 h^{mn}\partial_m\ln N\partial_n \ln N-{\cal K}\partial_{\hat{3}}\ln
 N\right.\nonumber\\-\left.\frac32{\cal K}^2+{\cal K}^{mn}{\cal K}_{mn}
+q^{mn}q_{mn} \label{26}
  \right).
\end{eqnarray}

Here the lapse is determined by the super-Hamiltonian constraint
\begin{eqnarray}\fl
  2\partial_m\!\left(|h|^\frac12 h^{mn}\partial_n \ln N\right)+
  2\partial_m\!\left(|h|^\frac12\theta^{\hat{3}m}{\cal K}\right)+|h|^\frac12
   \left(\frac12h^{mn}\partial_m\ln N\partial_n \ln N+2{\cal K}\theta^{\hat{3}m}
\partial_m\ln N\right.\nonumber\\\fl\left.-\frac32{\cal K}^2+{\cal K}^{mn}{\cal
K}_{mn}+q^{mn}q_{mn}\right)= 2\partial_m\left(|h|^\frac12
h^{mn}\partial_m\ln N\right) +|h|^\frac12
  \left(\frac12h^{mn}\partial_m\ln N\partial_n \ln N\right.\nonumber\\\left.
  +2{\cal K}
\partial_{\hat{3}}\ln N-2\partial_{\hat{3}}{\cal K}+\frac12{\cal
K}^2 \label{27}+{\cal K}^{mn}{\cal K}_{mn}+q^{mn}q_{mn}\right)=0.
\end{eqnarray}

Let us consider first of all the especially simple case of a maximal
 spacial Cauchy hypersurface  . Then the Hamiltonian
density (\ref{26}) takes the form
\begin{eqnarray}\fl
 {\cal H}^{SWOF}\left( {N} \right) = N|h|^\frac12\left({ {-\frac{{3}}{{2}}}}
 h^{mn}\partial_m\ln N\partial_n \ln N+{\cal K}^{mn}{\cal K}_{mn}
+q^{mn}q_{mn} \label{28}
  \right),
\end{eqnarray}
and will be everywhere positive definite if on $\Sigma$ exists an
appropriate solution of the super-Hamiltonian constraint
\begin{eqnarray}\fl
\label{29} 2\partial_m\left(|h|^\frac12 h^{mn}\partial_m\ln
N\right) +|h|^\frac12
  \left(\frac12h^{mn}\partial_m\ln N\partial_n \ln N+{\cal K}^{mn}{\cal K}_{mn}
+q^{mn}q_{mn}\right)=0.
\end{eqnarray}
Unique positive solution $N$ of this equation exists because
Nester's gauge (\ref{10}) enjoys the property of conformal
invariance and thus fits into the
Lichnerowicz--Choquet-Bruhat--York initial-value problem analysis
\cite{nesCQG}. Therefore, we conclude, that owing to the
correspondence between the SWE and the Nester gauge on the maximal
hypersurface there exists the direct relationship between the
Hamiltonian based positivity localization in the 4-covariant
spinor method and in the ADM method based on the SOF for the case
$N=\Phi^4$ .

Now, let us consider the   hypersurface $\Sigma$ which is not
maximal, and let be asymptotically $N=a+O(r^{-1}),\, \partial_m
N=O(r^{-2})$. Then the super-Hamiltonian constraint (\ref{27}) for
enough large $r$ can be written as
\begin{eqnarray}\label{30}\fl
2\partial_m\left(|h|^\frac12 h^{mn}\partial_m N\right)
+N|h|^\frac12
  \left(-2\partial_{\hat{3}}{\cal K}+\frac12{\cal K}^2
+{\cal K}^{mn}{\cal K}_{mn}+q^{mn}q_{mn}\right)=0.
\end{eqnarray}
The Dirichlet problem for equation (\ref{30})  has the unique
solution, if
\begin{eqnarray}\label{31}
  C(x)=|h|^\frac12\left(-2\partial_{\hat{3}}{\cal K}+\frac12{\cal K}^2
+{\cal K}^{mn}{\cal K}_{mn}+q^{mn}q_{mn}\right)\geq0.
\end{eqnarray}

The same condition, and the condition that $N$ is positive    on
the boundary or asymptotically,  ensure the non-occurence of the
knot points of equation (\ref{30}), since the knot submanifolds of
elliptic equation of second order are closed or their ends lie on
the boundary.  That is why everywhere $N>0$, and we can choose
$a=1$.

Further, a general theorem for the elliptic  second-order system
claims \cite{lop} that its solutions continuously depend on
coefficients, domain and values of the functions on the boundary,
therefore the Hamiltonian density $ {\cal H}^{SWOF}\left(N({\cal
K}), N \right)$ (\ref{26}) continuously depends on ${\cal K}$ and
thus is non-negative on the  hypersurfaces  which satisfy the
condition
\begin{eqnarray}\label{32}
-2\partial_{\hat{3}}{\cal K}+\frac12{\cal K}^2\geq0
\end{eqnarray}
and lie in some neighborhood of the maximal one. The presence of the
terms $-2\partial_{\hat{3}}{\cal K}$ and $\frac12{\cal K}^2$ in
the right side of relationship (\ref{31}) is caused just by a
fact that we used the SWOF; the application of Nester's gauge
does not give a possibility to prove the existence of this class
of hypersurfaces, on which the Hamiltonian density in the SOF is
non-negative.

In order to establish a correspondence between conditions
(\ref{11}) and (\ref{32}) we write following a space spinors
definition
\begin{eqnarray}\label{33}
  D_{A}{}^{B}{\cal K}=-\sqrt2
  n_{\hat{\alpha}}\sigma^{\hat{\alpha}}{}_{A\dot{A}}
  \sigma^{\hat{\beta}B\dot{A}}\partial_{\hat{\beta}}{\cal K}=
-\sqrt2
  \sigma^{\hat{0}}{}_{A\dot{A}}
  \sigma^{\hat{\beta}B\dot{A}}\partial_{\hat{\beta}}{\cal K},
\end{eqnarray}
and obtain that the diagonal elements of matrix
\[\frac{\sqrt2}{4}D_{A}{}^{B}{\cal K}+\frac12 \varepsilon
_A{}^B{\cal K}^2
\]
are
\[
\frac14\partial_{\hat{3}}{\cal K}+\frac12{\cal K}^2\qquad
\mbox{and} \qquad-\frac14\partial_{\hat{3}}{\cal K}+\frac12{\cal
K}^2.
\]
Therefore, the second of them is non-negative on the hypersurfaces
which satisfy condition (\ref{32}). This means that under
fulfilling of condition (\ref{32}) there is also fulfilled
condition b) of Theorem 3.

 So, if the SWE and conditions a) and c) of Theorem 3 are fulfilled, then
 on hypersurfaces, which satisfy condition  (\ref{32}) and lie in some
 neighborhood of the maximal one, the Hamiltonian density
${\cal H}\left( {\varphi ,\chi}  \right)$ (\ref{23}) and the ADM
Hamiltonian density ${\cal H}^{SWOF}\left( {N} \right)$ (\ref{26})
 are locally nonnegative simultaneously.

Let us note that just an absence of the result about connection
between the SWE equation and the Nester gauge (a theorem like
Theorem 3 and Theorem 4) did not permit  Nester and Tung to obtain
a direct relationship between the 4-spinor 3-form of the
Hamiltonian density under fulfilling of the SWE  and the
Hamiltonian density in the SOF formalism, both on the enough
general hypersurfaces and even on the maximal ones. The 3-spinor
formalism, developed by these authors
 and Zhytnikov, ensure  the partial solving  of this
problem; in particular, the energy is guaranteed to be locally non-negative
only on the maximal hypersurfaces.

\section{Conclusions}

Generalization  of the SOF by the SWOF allows us to remove two liabilities of
the SOF method: necessity of the restriction to the maximal hypersurfaces,
and impossibility of extension to the future null infinity and, hence,
description of the  Bondi 4-momentum. Therefore, it follows that  for  the
quasilocal Hamiltonian density (\ref{2}) investigation    are  suitable not
the Dirac equation and the 3-spinors, but  the SWE and the space spinors
introduced by Sommers \cite{som}. Although the 3-dimensional Dirac equation
and the SWE are very similar, we see that fixing of the spinor field by the
Dirac equation or by the SWE  leads to different physical consequences. The
mathematical consequences for application of these gauge conditions for the
spinor field are also different; in particular, the conditions for existence
of  solutions existence differ  in domains of  finite measure \cite{peljpa}.

The equivalence of the Sen-Witten spinor field and the SWOF
(Sec$.\, 2$), under the reasonable from the physical point of view
fulfilling of conditions of Theorem 3, permits to establish  that
the method of the 4-covariant quadratic spinor Hamiltonian and the
SOF method are very close.  The spinor parameterization of the
Hamiltonian displacement and correlations (\ref{14}) are a key for
 the orthonormal frame interpretation of the Hamiltonian 4-covariant
spinor form (\ref{2}) and the spinor interpretation of the ADM Hamiltonian
density even in the case when the spinor field or the orthonormal frame are
not fixed.

 Note at the end that conditions of (\ref{11}) and
(\ref{32}) type are the only sufficient ones, and we expect to
weaken then significantly or to exclude completely.

\section{Acknowledgment}
I wish to thank referees for helpful discussion.

\end{document}